\documentclass{ws-p8-50x6-00}

\def\lesim{\mathrel{\hbox{\rlap{\hbox{\lower4pt\hbox{$\sim$}}}\hbox{$<$}}}}
\def\gtsim{\mathrel{\hbox{\rlap{\hbox{\lower4pt\hbox{$\sim$}}}\hbox{$>$}}}}
\newcommand{\Deg}{{}^{\,\mbox{\scriptsize o}}}          % -  degree
\newcommand{\Exp}[1]{{\mbox{\Large \rm e}}^{\normalsize \it #1}}
\begin{document}

\title{Expected Performance of a Neutrino Telescope for
Seeing AGN/GC Behind a Mountain}

\author{George W.S. Hou and M.A. Huang}
\address{Department of Physics, 
	National Taiwan University, Taipei, Taiwan, R.O.C. \\
	E-mail: wshou@phys.ntu.edu.tw, huangmh@phys.ntu.edu.tw}

\maketitle

\abstracts{
We study the expected performance of building a neutrino telescope, which 
targets at energy greater than $10^{14}$ eV utilizing a mountain to interact 
with neutrinos. The telescope's efficiency in converting neutrinos into 
leptons is first examined. Then using a potential site on the Big Island of 
Hawaii, we estimate the acceptance of the proposed detector. The neutrino 
flux limit at event rate 0.3/year/half decade of energy is estimated to be 
comparable to that of AMANDA neutrino flux limit at above $10^{16}$ eV.
}

\section{Neutrino Astronomy}

Neutrino astronomy is still in its infancy. Although neutrinos are 
abundantly produced in stars, as they live and when they die, one suffers 
from an extremely low cross section for detection on Earth. Still, it is 
rather impressive that we already have ``neutrino images" of the Sun, as 
well as neutrino blips of the cataclysmic SN1987A event. At the start of a 
new century/millennium, we yearn to reach beyond the stars and observe 
cosmological neutrino sources. Large ``km$^3$" ice/water or air shower 
neutrino ``telescopes" are being built, and ``the sky is the limit".

Neutrinos could play an important role in connecting several branches of 
particle astrophysics. The origin of ultra-high energy cosmic rays (UHECR) 
is still a great puzzle~\cite{UHECR}. Bottom-up theories propose that they 
originate from energetic processes such as Active Galactic Nuclei (AGN) 
or Gamma Ray Bursts (GRB). The energetic hadron component could interact 
with accreting materials near the central black hole and produce 
neutrinos through the decay of charged pions. On the other hand, top-down 
theories suggest that UHECR are decay products of topological defects or 
heavy relic particles. According to these theories, there are more neutrinos 
than gamma rays and protons~\cite{Sigl}. Measurement of the neutrino flux at 
and above the ``knee" region (i.e. $\gtsim 10^{15}$ eV) provides a good 
discriminator to distinguish between the two scenarios. 

Cosmic gamma rays are attenuated by the infrared, microwave and radio 
background photons~\cite{gamma}.
%% length is in he order of kpc for PeV-$\gamma$ and Mpc for TeV-$\gamma$. 
The recent observation of TeV gamma rays from extragalactic sources such as 
Mkn421~\cite{Mkn421} and Mkn501~\cite{Mkn501}, however, 
has aroused some concern. In order to reach the Earth from extragalactic 
distances, these $\gamma$ sources must have either a much harder spectrum or 
more powerful mechanisms, e.g. electromagnetic (EM) processes such 
as inverse Compton scattering, or hadronic processes such as $p + X 
\rightarrow \pi^o + \ldots \rightarrow \gamma + \ldots$
The former produces few neutrinos, while the latter produces comparable 
amounts of both neutrinos and photons. Neutrinos therefore provide 
direct probes of the production mechanism of TeV $\gamma$ rays
from extragalactic sources such as AGN/GRB.

Recent results on atmospheric neutrinos add an interesting twist to 
cosmological neutrino detection. Super-Kamiokande (SK) and Sudbury Neutrino 
Observatory (SNO) data strongly suggest that {\it muon neutrinos oscillate 
into tau neutrinos}. Below $10^{12}$ eV, the tau decay length is less than 5 
mm, and SK and SNO have difficulty distinguishing between showers initiated 
by electrons and those by taus. Above $10^{15}$ eV, the tau decay length 
becomes 50 m or more, distinctive enough for identifying the taus. 
Since cosmic neutrinos are produced via $\pi^+$ decay predominately, 
one does not expect much directly produced cosmic $\nu_\tau$ flux 
\cite{Cheung}. Detecting a tau decendent on Earth would not only probe 
AGN/GRB mechanisms, but would also constitute a tau-appearance experiment.

\section{A Genuine Neutrino Telescope}

Because of the low interaction cross-section, all neutrino experiments 
resort to a huge target volume. The target volume is usually surrounded 
by the detection devices in order to maximize detection efficiency. Thus, 
the target volume is approximately equal to the detection volume. In other 
words, the cost of building a detector cost varies in propotional to the 
target/detection volume. Furthermore, to shield against cosmic rays or even 
high energy atmospheric neutrinos, these detectors often have to be deep 
underground. For instance, the km$^3$ size ICECUBE project \cite{Halzen} at 
the South Pole has a price tag of \$100M, aims to look for upward going 
events, and takes years to build. Variants such as sea/ocean or air watch 
experiments are similarly large and costly.
These ``telescopes" tend to bear litte resemblance to their EM counterparts.

Some alternative approaches have been proposed, such as using the 
Earth~\cite{Domokos} or a mountain~\cite{Fargion,Vannucci} as the target into 
convert neutrinos to leptons, which will then initiate air showers in the 
atmosphere. Observing the air showers from a region obscured by a mountain or 
the Earth can eliminate the contamination of cosmic ray showers. 
The main difference between this approach and the conventional
experiments is that the target volume and the detection volume are different. 
Moreover, materials in the target volume (mountain, Earth) and the detection 
volume (atmosphere) are readily available at almost no cost, thus the 
overall cost (and perhaps schedule) of the experiment could be reduced 
dramatically. This makes the approach worthy of further exploration.

\begin{figure}[tbp]
\epsfxsize=11cm
\epsfbox{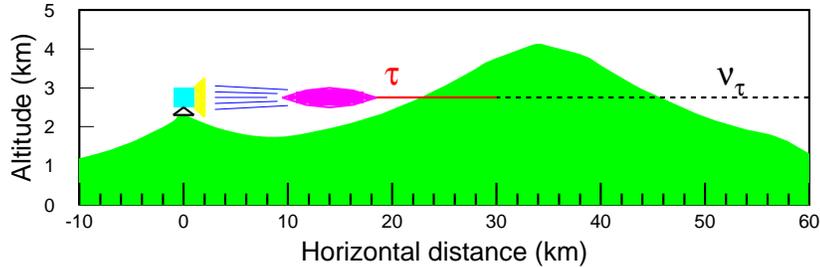} % postscript image file name
\caption{Illustration of a neutrino convert to tau inside a mountain. 
\label{fig:demo}}
\end{figure}

Using an approach similar to that of Vannucci~\cite{Vannucci}, a Cerenkov 
telescope sits on one side of a valley opposite a mountain. Energetic cosmic 
neutrinos, while passing through the atmosphere with ease, interact inside 
the mountain and produce leptons. Electrons will shower quickly and have 
little chance of escaping from the mountain. For muons, the 
decay/interaction lengths are too large to initiate showers inside the 
valley. The taus have suitable decay length to escape from the mountain 
{\it and} initiate showers inside the valley upon decay. 
This process is illustrated in Fig.~\ref{fig:demo}. 
With this design, 
the telescope is not only a detector for astrophysical and cosmological 
neutrinos, but also serves as a tau-appearance experiment.

It is interesting to note that this telescope resembles closely usual 
EM telescopes and a typical particle experiment. The field piece is the 
mountain, which functions as both a target and a shield,
and the subsequent valley is the shower volume.
The actual Cerenkov telescope functions as an ``eye piece" that focuses the 
Cerenkov light from a shower emerging from the mountain onto a sensor plane.
The sensor could be a MAPMT array, where fast electronics matches the 10 
ns Cerenkov pulse and helps discriminate against other background sources.
Using two telescopes in coincidence would be produce better results.
The only drawback, in comparison to a regular EM telescope, is that we cannot 
move the mountain and would have to rely on the Earth's rotation to move the 
telescope.

Besides cost, expected to be far less than ICECUBE or Auger,
the most critical issue is the expected {\it count rate}.
In the following, we choose a potential site (Hawaii Big Island),
examine the neutrino conversion efficiency, and then derive the
flux limit and sky coverage of the proposed detector.

\section{Potential Site}

The criteria for choosing a potential site are as follows:
\begin{itemize}
\item Reduced artificial lights, dry air and cloudless sky, much like usual 
optical telescopes.
\item Target mountain broad enough for the sake of acceptance.
\item Valley wide enough for taus to decay and air showers to develop.

In the energy range of $10^{14}$ -- $10^{18}$  eV, the depth 
of shower maximum $\sim$ 500 -- 800 gm/cm$^2$. 
At altitude around 2 km, this depth corresponds to a horizontal distance of 
4.5 to 7.8 km. Therefore, the width of the valley must be 
larger than 5 km, but less than the attenuation length of light $\sim$ 50 km.
\item Good exposure to the Galaxy Center (GC).

The nearest massive black hole --- what may be behind astrophysical 
neutrinos --- is our Galaxy Center.
\end{itemize}

Hawaii Big Island, with its perfect weather conditions, has been a favorable 
site for astronomical (EM) telescopes. 
The Big Island also has a rather unique configuration. Besides the more 
sought after Mauna Kea, the other 4 km high mountain, Mauna Loa, has a 
breadth of approximately 90 km. 
Across from Mauna Loa to the northwest, Mount Hualalai is $\sim$ 20 
km away and 2.3 km in altitude. This makes Mauna Loa a good candidate for 
the target mountain with the detector installed on top of Hualalai. 
In the following study we assume this configuration.

\section{Neutrino conversion efficiency}

In this study, the mountain is simplified as a block of thickness L. 
Neutrinos enter the mountain, pass through distance $x$, interact in $x$ to 
$x+dx$, produce taus, which then survive through the rest of the mountain 
without decay. 

The probability for neutrinos to survive the atmosphere ($P_1$) is 
taken as 1, which is very close to the actual case.
%
%$P_1(X_{atm})=\exp(-X_{atm}/\lambda_{\nu})$, 
%where $X_{atm}$ is the depth of atmosphere and $\lambda_{\nu}$ is the 
%interaction depth. $\lambda_{\nu}=1/(N_A\times\sigma)$, where
%$\sigma$ is the charged current interaction 
%cross-section \cite{Gandhi} and $N_A$ is the Avogadro number. This 
%probability 
%is irrelevant to the conversion inside a mountain, we include it for the 
%completeness of physics. $X_{atm}$ is assumed to be 3000 $gm/cm^2$ for 
%zenith angle of $89.5\Deg$.
%
The probability for neutrinos to survive distance x inside the mountain is 
$P_2(X) = \exp(-x / \lambda_{\nu})$, where $\lambda_{\nu}=1/(N_A 
\sigma\rho)$, $\sigma$ is the charged current interaction 
cross-section \cite{Gandhi}, $N_A$ is the Avogadro number, and $\rho$ is 
the mean density of the mountain. 
The chance of neutrino interaction in $x$ to $x+dx$ is $dx/\lambda_{\nu}$. 
The energy of tau is approximated as $E_{\tau}= (1-y) E_{\nu}$, where $y$ is 
the fraction of energy carried by the recoiling (shattered) nuclei or 
electron, which is in the range of 0.2 to 0.5 with mean $\sim$ 0.25, we 
therefore use $E_{\tau}= 0.75 E_{\nu}$.

The probability for taus to survive through the rest of the mountain of 
distance $L-x$ is $P_3(X) = \exp(-(L-x) / \lambda_{\tau})$, where 
$\lambda_{\tau}$ is the decay length of tau and equals 
$(E_{\tau}/{\rm PeV})\; 48.91$ m. 

\begin{figure}[htbp]
%\figurebox{20pc}{11cm}{} % to have a box alone
%\epsfig{figure=eff_wid_e.eps,width=11cm} % postscript image file name
 \epsfxsize=11cm 
\epsfbox{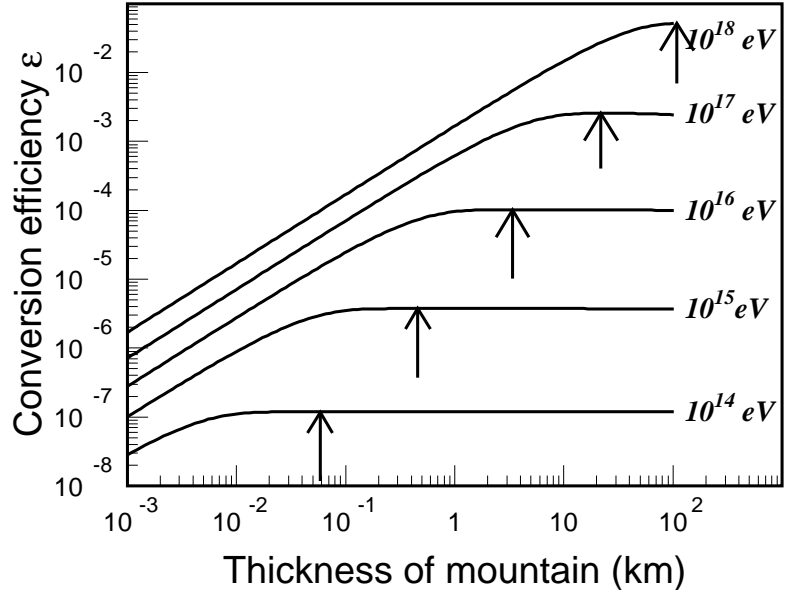} % postscript image file name
\caption{Neutrino conversion efficiency vs mountain thickness (in kilometers) 
for five energies. The maximum efficiencies are marked with arrows.
\label{fig:eff_wid}}
\end{figure}

The neutrino conversion efficiency is 
\begin{equation}
%% &=& \int_0^L P_2(x)P_3(L-x) \frac{dx}{\lambda_{\nu}} \nonumber \\
\varepsilon = \int_0^L \Exp{-x/\lambda_{\nu}} \Exp{-(L-x)/\lambda_{\tau}} 
		\frac{dx}{\lambda_{\nu}}
            =  \frac{\lambda_{\tau}}{\lambda_{\nu} - \lambda_{\tau}} 
	\left( \Exp{-L/ \lambda_{\nu}} - \Exp{-L/ \lambda_{\tau}} \right), 
\label{eq:convert}
\end{equation}
where the integration was done by {\it neglecting the energy loss of tau}.
The maximum efficiency occurs at 
$
\frac{d\varepsilon}{dL} | _{L=L_{max}}=0,
$
%\[ \frac{\Exp{-L/\lambda_{\nu}}}{ \lambda_{\nu}} =
%	\frac{\Exp{-L/\lambda_{\tau}}}{ \lambda_{\tau}} \]
i.e.
\begin{equation} 
L_{max} = \frac{\ln(\lambda_{\nu}/\lambda_{\tau})}
	{\frac{1}{\lambda_{\tau}} - \frac{1}{\lambda_{\nu}}} 
\end{equation}
The conversion efficiencies at five energies are shown in 
Fig.~\ref{fig:eff_wid}. The efficiency  plateaus above $L > L_{max}/2$.
This maximal efficiency scales roughly as a power law in energy.
\[ \varepsilon_{max} \simeq \frac{\lambda_{\tau}}{\lambda_{\nu}}
	\simeq 3.6\times 10^{-6} (\frac{E_{\nu}}{\mbox{PeV}})^{1.4} \]
%at $L_{max}$ as a function of energy are 
%shown in Fig.~\ref{fig:eff_eng}.
%From Fig. 1, in consideration of a typical mountain,
%one would start to lose sensitivity above $10^{17}$ eV.

%\begin{figure}[htbp]
%\figurebox{20pc}{11cm}{} % to have a box alone
% \epsfxsize=11cm % will enlarge or reduce the postscript figures based on 
%the xsize
%\epsfig{figure=effmax_e.eps,width=11cm} % postscript image file name
%\caption{The maximal neutrino conversion efficiency as a function energy.  
%\label{fig:eff_eng}}
%\end{figure}

The mean distance traveled by taus inside the mountain is
\begin{equation}
\overline{L_{\tau}} = \frac{\int_0^L (L-x) \Exp{-x/\lambda_{\nu}} 
\Exp{-(L-x)/\lambda_{\tau}} \frac{dx}{\lambda_{\nu}}}
	{\int_0^L \Exp{-x/\lambda_{\nu}} \Exp{-(L-x)/\lambda_{\tau}} 
	\frac{dx}{\lambda_{\nu}}} \nonumber \\
%	&=& \frac{1}{\frac{1}{\lambda_{\tau}} -  \frac{1}{\lambda_{\nu}}} 
= \frac{\lambda_{\nu}}{\lambda_{\nu} - \lambda_{\tau}} \lambda_{\tau}.
\end{equation}
Because $\lambda_{\nu}\gg \lambda_{\tau}$, 
$\overline{L_{\tau}}\lesim \lambda_{\tau}$, the mean production point of
 tau is approximately one decay length inside the mountain. As long as 
the thickness of mountain is larger than $\lambda_{\tau}$, 
$\overline{L_{\tau}}$ remains unchanged.

\section{Acceptance of flux limit}

\begin{figure}[htbp]
 \epsfxsize=11cm 
\epsfbox{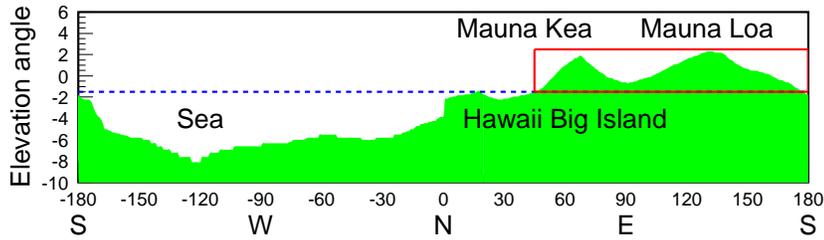} % postscript image file name
\caption{The panoramic view from the top of Hualalai. The dash line is the 
horizon and the shaded region is the field of view obstructed by the terrain 
of Hawaii Big Island. The region between the horizon and the terrain is the 
sea to the west of the Big Island. 
\label{fig:fov}}
\end{figure}

\begin{figure}[htbp]
 \epsfxsize=11cm
\epsfbox{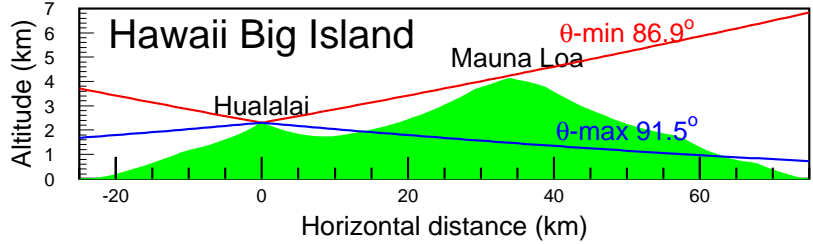} % postscript image file name
\caption{A cross-section of Big Island along the line from 
Hualalai to Mauna Loa.
\label{fig:theta}}
\end{figure}

Fig.~\ref{fig:fov} shows the panoramic view from the top of Mt. Hualalai 
towards Mauna Loa. The field of view of the detector is the shaded mountain 
region inside the box. The azimuth angle extends from south to east. 
The minimum zenith angle of $86.9\Deg$ is set by the line from the summit of 
Hualalai to that of Mauna Loa. The maximum zenith angle of $91.5\Deg$ is set 
by the line from the summit of Hualalai to the horizon at the base of Mauna 
Loa. A cross-section of the Big Island along the line from 
Hualalai to Mauna Loa is shown in Fig.~\ref{fig:theta}.

The acceptance is defined by the effective area multiplied by the effective 
solid angle. 
Owing to lateral distribution of air shower, the Cerenkov light cone of 
shower is % much wider. This angle is 
approximately $5\Deg$ to $6\Deg$~\cite{PS}. The effective solid angle can
be determined by the sensitivity of PMT, the distance from the detector to the 
shower maximum, and the Cerenkov light yield of air shower. 
The number of Cerenkov photons is proportional to the number of secondary 
particles in the air showers, which is proportional to the tau energy. 
Also, the lower energy taus decay closer to 
the mountain, thus farther away from the detector and the Cerenkov 
light suffers more atmospheric scattering. These two effects 
reduce the effective solid angle at lower energy. 
The extend of the effect can be obtained by detailed simulation. 
To simplify the calculation, we use a 
constant value of $5\Deg$, which yields the effective solid angle 
\[ \Omega = \int_0^{\theta_c} \sin\theta d\theta d\phi 
	= 2\pi\,(1-\cos\theta_c) = 0.024\; \mbox{sr}	\] 

\begin{figure}[htbp]
\epsfxsize=11cm 
\epsfbox{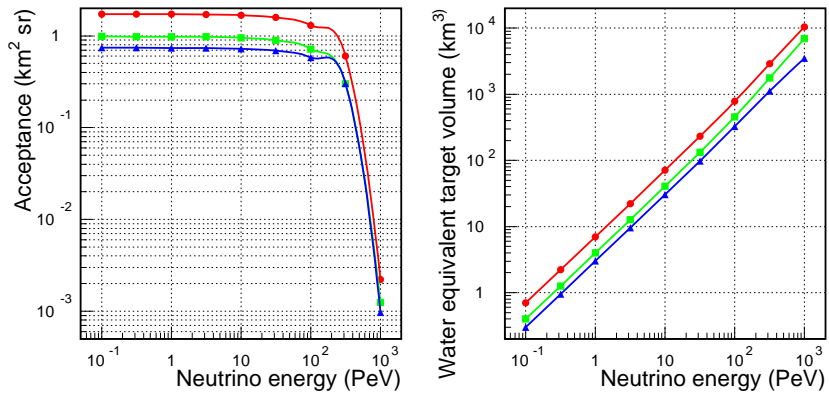} % postscript image file name
\caption{Acceptance and water equivalent target volume of the potential 
site in Hualalai. 
\label{fig:acc_eng}}
\end{figure}

The effective area is the cross-section area where tau decays. The mean 
distance of decay after taus escape from the mountain is still 
$\lambda_{\tau}$. So the effective area is
\[ 
a_{\rm eff}(E) =\int_{\rm FOV} (r(\omega) - \lambda_{\tau}(E))^2 d\omega 
\]
where $\omega$ is the solid angle of each pixel, FOV is the field of view, 
and $r$ is the distance from the detector to the mountain surface viewed by 
that pixel. The total acceptance $A(E)$ is $a_{\rm eff}(E) \times \Omega$ 
as shown in Fig.~\ref{fig:acc_eng}.
Below $10^{17}$ eV, the acceptance is approximately 1 km$^2$ sr, 
similar to ICECUBE. The sharp decrease in acceptance at 
$E> 2\times 10^{17}$ eV is due to the increase in decay length of tau 
beyond 10 km. The valley $\sim$ 20 km is not wide enough to contain these 
high energy taus.

The target volume is defined as the volume inside the mountain where taus 
are produced,
\[	V = \int_{\rm FOV} \int_{R_i}^{R_f}  r(\omega)^2  dr d\omega, 	\]
where $R_i$ is the distance from the detector to the mountain surface, and
\[ 
R_f = \left\{ \begin{array}{ll}
	R_i + W & \mbox{if}\ W< \overline{L_{\tau}} \\
 	R_i + \overline{L_{\tau}} & \mbox{if}\ W \ge \overline{L_{\tau}} 
	\end{array} \right.
\]
where $W$ is the width of mountain in the field of view $\omega$. The 
target volume is then transferred to the water-equivalent target volume by 
multiplying the density of rock, $2.65g/cm^3$. 
$\overline{L_{\tau}} \simeq \lambda_{\tau}$ increases almost linearly with 
energy, so does the target volume. For conventional neutrino telescopes, 
such as SK or ICECUBE, the target volume is identical to the detection 
volume, 
therefore the acceptance is propotional to the detection volume.
For the Earth-skimming or mountain-valley type neutrino telescopes, 
the target volume and the detection volume are different. Thus, the 
acceptance and the target volume do not have any direct relation.

This study does not consider the effect of energy loss of taus inside the 
mountain. The effect becomes more serious for energies $> 10^{17}$ eV, 
where tau energy loss leads to a decrease in decay length of taus, 
thus {\it increasing} acceptance at high energy.
When the energy loss of tau is taken into consideration, 
Eq.~(\ref{eq:convert}) cannot be integrated in closed form. 
At the present stage, we have ignored the energy loss 
effect for simplicity and treat the results as lower limit of acceptance 
and upper limit of sensitivity. 

Because of the lower light yield and more scattering at lower energy, the 
acceptance should be lower at lower energy.
In view of the two factors above, the best energy range for this type of 
detector is approximately in $10^{15} < E < 10^{18}$ eV.

The flux limit is estimated by 
\[ \Phi(E)=\frac{d^2N(E)}{dT \;dE} = \frac{dN(E)/dT}{A(E) \varepsilon(E)} \]
where $N$ is the number of events, $T$ is the exposure time, $dN(E)/dT$ is 
the event rate, $dE$ is the bin width of energy which is approximately equal 
to the energy resolution of detector. The conversion efficiency
$\varepsilon(E)$ is calculated by similar process as Eq.~(\ref{eq:convert}). 
The exact zenith angle, the atmospheric pressure, the mountain width, and the 
curvature of the Earth are all taken into consideration.

In the conversion from $\nu_{\tau}$ to $\tau$, some fraction of energy ($y$E)
are brough out by interacting nuclei. Because this interaction take place 
inside the mountain, this energy can not be measured.  
$\sigma_y \sim 0.18$ is the largest source of systematic error in energy. 
With some uncertainties from detection and reconstruction, a 
simplified value of half a decade $10^{-0.5}=0.31$ is {\it assumed} as the 
energy resolution $dE$.

\begin{figure}[htbp]
\epsfxsize=11cm
\epsfbox{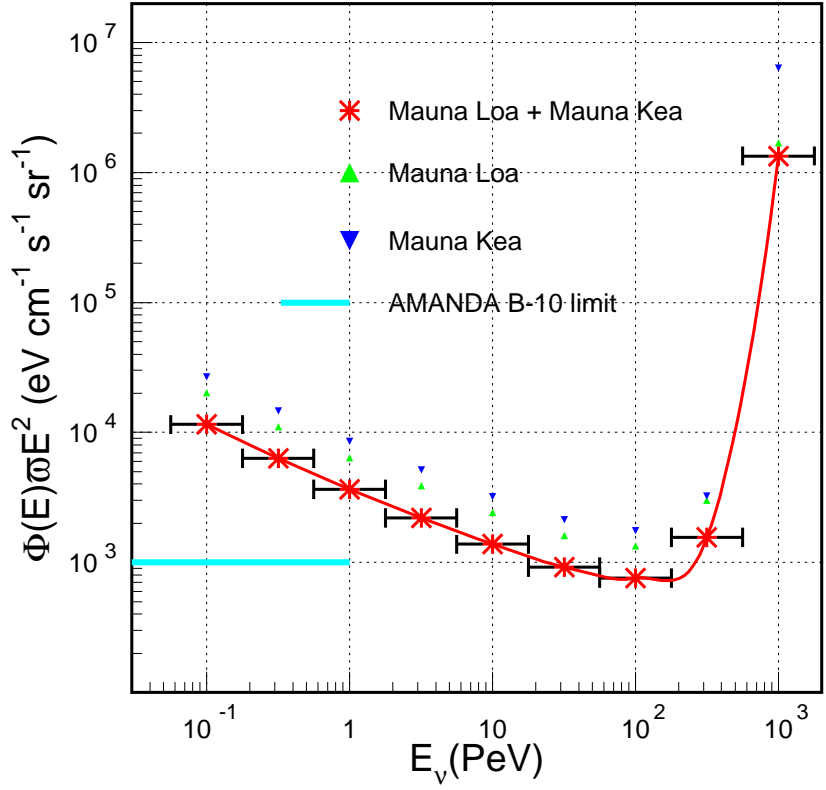} % postscript image file name
\caption{The sensitivity of the proposed detector if the event rate is 0.3 
event/year/half decade of energy.
\label{fig:limit}}
\end{figure}

The detector sensitivity is defined as the flux when the event rate is 0.3 
event in one year. Based on the acceptance of Hualalai site, the sensitivity 
of the proposed detector and the recent AMANDA B-10 neutrino limit are shown 
in Fig. \ref{fig:limit}. Note that the AMANDA B-10 limit is the integral flux 
limit from null observation of neutrino in the energy range of $10^{12}$ to 
$10^{15}$ eV. The null observation in one year of operation of the 
proposed detector could set an upper limit similar to that of 
AMANDA B-10~\cite{AMANDA}, but at $10^{15} < E < 10^{18}$ eV. 

\section{Sky coverage}

The detector is operated at moonless and cloudless nights.  
We simulate the operation from December 2003 to December 2006. The detector 
operates when the total time of moonless night is longer than one hour. The 
total exposure time in three years is 5200 hours, corresponding to a duty 
cycle of $\sim$ 20\%. In reality, some cloudy nights have to be excluded. 

\begin{figure}[htbp]
\epsfxsize=11cm
\epsfbox{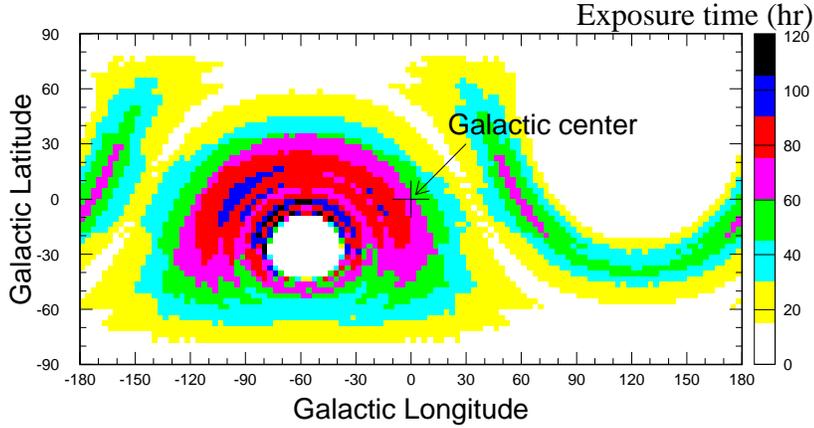} % postscript image file name
\caption{The exposure time in galactic coordinates for the three-year 
operation from December 2003 to December 2006.
\label{fig:gal}}
\end{figure}

According to the field of view specified above, the sky covered by the 
detector can be calculated. The total exposure hours in $1\Deg \times 1\Deg$
of galactic coordinates are shown in Fig. \ref{fig:gal}. The galactic center 
is visible for approximately 70 hours.

\section{Discussion}
Although the acceptance reaches 1 km$^2$ sr, the optical detection suffers 
10\%  operation time in each calandar year. There are several ways to improve 
the acceptance.
\begin{itemize}
\item Extending the zenith angle coverage to below the horizon can include 
the 
	Earth-skimming events, which are not studied in this report. 
	This extension could double the acceptance at $E<10^{16}$ eV. At 
	higher energy, acceptance does not increase much because of lack of 
	space for taus to decay.
\item If the detector could also detect the fluorescent light from air 
	showers, the current field of view could be 
	triggered by showers initiated by the taus escaping from Mauna Kea and 
	by Earth-skimming from south-west of Mauna Loa. The large increase in 
	solid angle could increase acceptance by a factor of 3 to 10. This is 
	most effective at energy higher than $10^{17}$ eV.
\end{itemize}
The above improvements can increase the acceptance to 20 km$^2$ sr. 
The azimuth angle can also be extended to the west 
side of Hualalai so that the sea-skimming events can be used as well. 
However, the reflection from waves may create more noise. 

The detector should have some coverage of the sky and record cosmic ray 
events. This can help monitor detector performance, and 
the cosmic ray flux can be used to cross-calibrate the energy scale with 
other 
cosmic ray experiments.

\section{Summary}

Taking Hawaii Big Island as a potential neutrino telescope site, we calculate
the neutrino conversion efficiency. The detector acceptance is
approximately 1.4 km$^2$ sr. The sensitivity of the proposed detector is 
close to the AMANDA B-10 limit. The exposure time of the galactic center, 
where the nearest black hole is located, is approximately 70 hours in three 
years of operation. 

This study shows that a compact neutrino telescope utilizing the 
mountain for neutrino conversion is capable of achieving a sensitivity 
similar to that of a big detector. In addition, the cost and construction 
time is greatly reduced. This type of detector at $10^{15} < E < 10^{18}$ eV 
could complement conventional neutrino telescopes such as 
AMANDA aiming at energies $E \lesim 10^{16}$ eV, and 
cosmic ray experiments such as Auger aiming at $E \gtsim 10^{18}$ eV.

\section*{Acknowledgments}

The authors would like to thank the HiRes group for providing the source 
code for moonless nights in Julian time.

\end{document}